\documentclass[11pt,a4paper,english]{elsart}
\usepackage[T1]{fontenc}
\usepackage[latin1]{inputenc}
\usepackage{amsmath}
\usepackage{graphicx}
\usepackage{amssymb}

\makeatletter

\providecommand{\tabularnewline}{\\}

\allowdisplaybreaks

\usepackage{bm}
\usepackage{slashed}
\usepackage{url}
\usepackage{braket}

\newcommand{\sst}[1]{\scriptstyle{#1}}
\newcommand{\ssst}[1]{\scriptscriptstyle{#1}}

\newcommand{\PV}{\slashed{P}}
\newcommand{\HPV}{H_{\ssst{\slashed{P}}}}
\newcommand{\TV}{\slashed{T}}
\newcommand{\PVTV}{\slashed{P}\slashed{T}}
\newcommand{\HPVTV}{H_{\ssst{\slashed{P}\slashed{T}}}}

\usepackage{babel}
\makeatother
\begin{document}
\begin{frontmatter}

\title{Time-Reversal Violation in Threshold $\vec{n}$ $\vec{p}$ Scattering}

\author{C.-P. Liu, R.G.E. Timmermans}

\address{Theory Group, Kernfysisch Versneller Instituut, University of Groningen,
Zernikelaan 25, NL-9747 AA Groningen, The Netherlands}

\begin{abstract}
We investigate parity and time-reversal violation in neutron-proton
scattering in the optical regime. We calculate the neutron spin rotation
and analyzing power in scattering on polarized protons. This allows
us to quantify the sensitivity that such experiments should aim for
in order to be competitive to present-day measurements of the neutron
electric dipole moment in constraining the $P$- and $T$-odd two-nucleon
interaction. While state-of-the-art techniques fall short by some
three orders of magnitude for the neutron-proton case, specific neutron-nucleus
experiments look promising, provided certain experimental and theoretical
challenges are met.
\end{abstract}
\end{frontmatter}
The neutron is an excellent laboratory for the study of fundamental
symmetries and interactions. Its lifetime can be used to determine
$V_{ud}$, one of the Cabbibo-Kobayashi-Maskawa matrix elements~\cite{Eidelman:2004wy}.
The correlations between the various momenta and spins in neutron
$\beta$-decay are sensitive probes of non-($V-A$) currents~\cite{Eidelman:2004wy}.
The photon asymmetry $A_{\gamma}$ associated with radiative capture
of polarized neutrons by nuclei, and the spin rotation $\phi_{\textrm{spin}}$
picked up by polarized neutrons traversing through a medium, can be
used to constrain the strangeness-conserving, hadronic weak interaction
(see, \emph{e.g.}, Refs.~\cite{Adelberger:1985ik,Haeberli:1995}
for reviews). The results of these measurements provide important
tests of the electroweak sector of the Standard Model, and in particular
its aspect of parity violation ($\PV$).

Neutrons can play an equally important, and in some sense even more
fundamental, role in the aspect of time-reversal violation ($\TV$).
Because of $CPT$ invariance, $T$ violation~\cite{Wolfenstein:1999re}
is equivalent to $CP$ violation, whose origin and role in generating
the matter-antimatter asymmetry of the Universe are among the great
mysteries of particle and astroparticle physics. The search for a
permanent neutron electric dipole moment (EDM), which violates both
$P$ and $T$ invariance ($\PVTV$), has been continuously in the
spotlight~\cite{Eidelman:2004wy,Harris:1999jx}. Possibilities to
identify $\TV$ in nuclear $\beta$-decay or in neutron-nucleus interactions
have also been seriously considered (see, \emph{e.g.}, Refs.~\cite{Gudkov:1991qg,Boehm:1995}
for reviews). The study of this report falls into the latter category.

Modern high-flux, continuous or pulsed, neutron sources are able to
provide neutrons over a wide energy spectrum, ranging from very fast
($\gtrsim\textrm{MeV}$) neutrons all the way down to ultra-cold ($\lesssim10^{-7}\,\textrm{eV}$)
neutrons. For the study of the $\PV$ or $\TV$ hadronic interaction,
low-energy neutrons, from the epithermal ($\sim\textrm{eV}$) to the
cold ($\sim\textrm{meV}$) region, are particularly useful for several
reasons: ($i$) The large flux can be maintained. ($ii$) Because
of the long de Broglie wave-length of the neutrons, the scatterers
contribute coherently. In other words, in this energy regime {}``neutron
optics'' works well. ($iii$) Low-energy neutrons are better suited
to study the short-ranged $\TV$ hadronic interaction than charged
particles, which are kept apart by the repulsive Coulomb force.

The Spallation Neutron Source, which is currently under construction
at Oak Ridge National Laboratory, is expected to improve fundamental
neutron physics to a new level. For example, a proposal to measure
the $\PV$ neutron spin rotation in para-hydrogen (with unpaired proton
spins) is aiming to reach an accuracy of $\sim2.7\times10^{-7}\,\textrm{rad/m}$~\cite{Markov:2002}.
Motivated by this remarkable advance, we investigate here $T$ violation
in scattering of polarized neutrons ($\vec{n}\,$) on polarized protons
($\vec{p}\,$), for which the $\TV$ signal can be calculated reliably
by using modern high-quality strong $n\, p$ potentials together with
the general $\PV$ and $\TV$ interaction. The observables that we
are interested in violate both $P$ and $T$, and hence they address
the same physics as the neutron EDM, $d_{n}$ (or the EDM of a diamagnetic
atom, such as $^{199}$Hg~\cite{Romalis:2000mg}). Our main purpose,
in fact, is to quantify how such a neutron-optics experiment, now
with a polarized target but assuming the same experimental accuracy,
competes with modern EDM measurements in constraining the underlying
$\PVTV$ interaction.~%
\footnote{A $\TV$ interaction which conserves $P$ does not belong to the same
class as an interaction which generates EDMs, and therefore will not
be considered here.%
} Also, a number of studies indicate that $\PV$ observables can be
greatly enhanced in certain neutron-nucleus scattering processes (see,
\emph{e.g.}, Refs.~\cite{Bunakov:1982is,Soderstrum:1988,Shimizu:1992qh,Haseyama:2001mg}).
We will use these results to justify some reasonable assumptions that
will allow us to extrapolate our results from the $\vec{n}\,\vec{p}$
system to $T$ violation in neutron-nucleus scattering. Our calculations
can thus serve as a benchmark for gauging the sensitivity of $\TV$
observables in neutron transmission experiments that aim to compete
with EDM measurements.

The optics of low-energy neutron transmission through a medium (see,
\emph{e.g.}, Refs.~\cite{Hughes:1954,Gurevich:1968}) can be described
by the corresponding index of refraction, $n$, which is a coherent
sum of individual scatterings and which is related to the neutron-target
scattering amplitude at forward angle ($\theta=0$), $f$, by \begin{equation}
n=1-2\,\pi\, N\, f/k^{2}\,,\end{equation}
 where $N$ is the target density; $k\equiv|\bm k|$ is the neutron
momentum, which is assumed to be in the $+z$ direction from now on.
When $f$ contains some non-vanishing component $f_{\PV}$ which depends
on $\bm\sigma\cdot\bm k$ due to $\PV$ interactions, neutrons with
a $+z$ polarization have a different value of $n$ compared to the
ones with a $-z$ polarization. Neutron wave functions of opposite
polarizations then pick up different phases, \emph{viz.} $n_{+z}\, k\, l$
and $n_{-z}\, k\, l$, after travelling a distance of $l$ in a uniform
medium. This optical dichroism manifests itself in two major ways:
($i$) a neutron spin rotation $\phi_{z}$ along the $z$-axis, and
($ii$) a longitudinal polarization $P_{z}$ of an unpolarized incident
beam or a longitudinal asymmetry $A_{z}$ between $+z$- and $-z$-polarized
neutrons~\cite{Michel:1964,Stodolsky:1974hm,Stodolsky:1981vn}. The
former depends on the real part of $f$, while the latter on the imaginary
part, as \begin{align}
\phi_{z} & =-2\,\pi\, l/k\, N\,\textrm{Re}(f_{+z}-f_{-z})\,,\label{eq:phiPV}\\
P_{z} & =-2\,\pi\, l/k\, N\,\textrm{Im}(f_{+z}-f_{-z})\,.\label{eq:PPV}\end{align}

These ideas for $P$ violation were generalized to study $T$ violation
by Kabir~\cite{Kabir:1981tp} and Stodolsky~ \cite{Stodolsky:1981vn}.
With a polarized target (with polarization $\bm S$), the scattering
amplitude can acquire, in principle, a $\PVTV$ component $f_{\PVTV}$
proportional to the triple correlation $\bm\sigma\cdot\bm k\times\bm S$.
Bunakov and Gudkov, however, argued later~\cite{Bunakov:1984} that
the combined actions of the magnetic interaction, which introduces
a $\bm\sigma\cdot\bm S$-dependent component $f_{M}$ in $f$, and
the weak interaction, generate a much larger scattering amplitude
of the same $\bm\sigma\cdot\bm k\times\bm S$ form. This effect mimics
$T$ violation -- similar to how final-state interactions can mimic
the $\PVTV$ correlation coefficient $R$ in $\beta$-decay. Such
a pseudo-$\PVTV$ amplitude ultimately spoils the unambiguous identification
of a true $\PVTV$ signal. Several ways to circumvent this difficulty
have been proposed in Refs.~\cite{Stodolsky:1986,Kabir:1987tu,Kabir:1988ma}.
Here, we analyze two observables and show what they can reveal about
the underlying $\PVTV$ nucleon-nucleon ($N\! N$) interaction.

\begin{figure}

\caption{(a): A $90^{\circ}$ spin rotation of a $x$-polarized neutron around
the $y$-axis (perpendicular to the plane) when travelling through
a $x$-polarized target. (b): Time reversal of (a). (c): A $180^{\circ}$
rotation around the $y$-axis of (b) which shows a $-90^{\circ}$
spin rotation instead. Therefore, the combined $T$- and $R$-invariance
require a zero spin rotation along the $y$-axis. \label{fig:phiy} }

\includegraphics{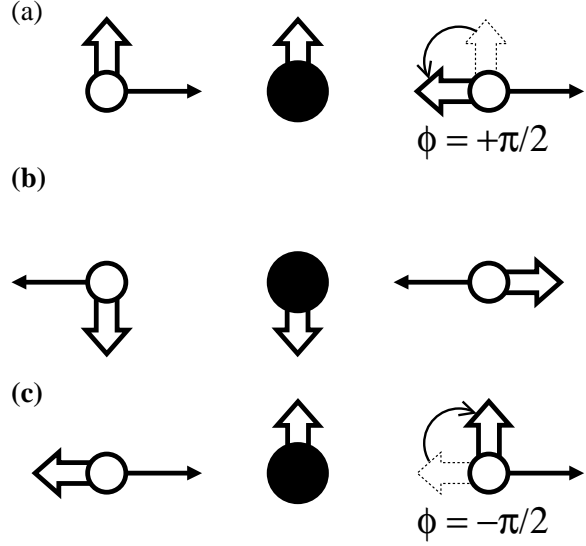}
\end{figure}

\begin{figure}

\caption{(a): The polarization $P_{y}$ of an unpolarized neutron beam when
travelling through a x-polarized target. (b): Time reversal of (a).
(c): A $180^{\circ}$ rotation along the y-axis of (b) which shows
the definition of asymmetry but with an additional minus sign $-A_{y}$.
Therefore, the combined $T$- and $R$-invariance require $P_{y}+A_{y}=0$.
\label{fig:py}}

\includegraphics{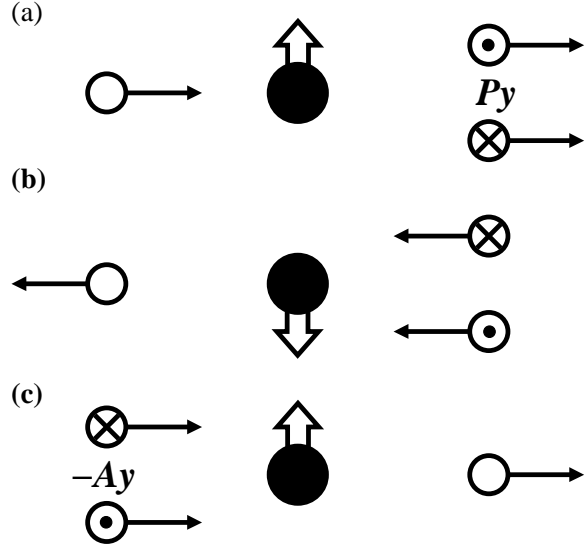}
\end{figure}

Without loss of generality, we assume that both neutron and proton
are polarized in the $+x$ direction. Because $\bm k\times\bm S$
defines a specific direction ($+y$ in our case), similar to $\bm k$
for the above $\PV$ case, the quantities $\widetilde{\phi}_{y}$,
$\widetilde{P}_{y}$, and $\widetilde{A}_{y}$ can be obtained via
the scattering amplitudes $\widetilde{f}_{+y}$ and $\widetilde{f}_{-y}$:\begin{align}
\widetilde{\phi}_{y} & =-2\,\pi\, l/k\,\widetilde{N}\,\textrm{Re}(\widetilde{f}_{+y}-\widetilde{f}_{-y})\,,\label{eq:phiPVTV}\\
\widetilde{P}_{y} & =-2\,\pi\, l/k\,\widetilde{N}\,\textrm{Im}(\widetilde{f}_{+y}-\widetilde{f}_{-y})\,.\label{eq:PPVTV}\end{align}
 We use here tildes as a reminder that we consider the case which
involves polarized targets and that it is the observables which violate
not only $P$ but also $T$ that are of interest. Analogously to what
has been concluded in Ref.~\cite{Kabir:1988ma}, one finds that ($i$)
$\widetilde{\phi}_{y}$ and ($ii$) $\widetilde{P}_{y}+\widetilde{A}_{y}$
are unambiguous measures of $T$ violation. This can be easily illustrated
in Figs.~\ref{fig:phiy} and~\ref{fig:py}: Although pseudo-effects
can mimic true $\TV$ effects in the scattering amplitude and some
observables, their invariances under $T$ and $R_{y}(\pi)$, a $180^{\circ}$
rotation around the $y$-axis, will render that \begin{align*}
\widetilde{\phi}_{y}^{\textrm{pseudo}} & =R_{y}(\pi)\, T\,\widetilde{\phi}_{y}^{\textrm{pseudo}}\, T^{-1}\, R_{y}^{-1}(\pi)=-\widetilde{\phi}_{y}^{\textrm{pseudo}}\,,\\
\widetilde{A}_{y}^{\textrm{pseudo}} & =R_{y}(\pi)\, T\,\widetilde{A}_{y}^{\textrm{pseudo}}\, T^{-1}R_{y}^{-1}(\pi)=-\widetilde{P}_{y}^{\textrm{pseudo}}\,.\end{align*}
 Therefore, neither ($i$) nor ($ii$) can be faked by a pseudo-effect.
It is also worth to point out that only one experiment is needed for
measuring $\widetilde{\phi}_{y}$, but two are needed for the $\widetilde{P}_{y}+\widetilde{A}_{y}$
comparison. In other words, the spin rotation represents a true null
experiment to test $\TV$, and therefore has some advantage~\cite{Arash:1985dy}.

The calculations of $f_{\PV}$ and $\widetilde{f}_{\PVTV}$ are briefly
outlined in the following. Since both $\PV$ and $\PVTV$ interactions,
$H_{\PV}$ and $H_{\PVTV}$, are much smaller than the strong interaction,
the first-order Born approximation is sufficient to calculate the
scattering amplitudes. Resolving the spin states for both neutron
and proton explicitly in terms of spinors quantized in the $z$-direction,
one obtains\begin{align}
f_{+z}-f_{-z} & =1/2\,\{\HPV(\uparrow\uparrow,\uparrow\uparrow)+\HPV(\uparrow\downarrow,\uparrow\downarrow)-\HPV(\downarrow\uparrow,\downarrow\uparrow)-\HPV(\downarrow\downarrow,\downarrow\downarrow)\}\,,\label{eq:Df-PV}\\
\widetilde{f}_{+y}-\widetilde{f}_{-y} & =-i/\sqrt{2}\,\{\HPVTV(\uparrow\uparrow,\uparrow\uparrow)+\HPVTV(\uparrow\downarrow,\uparrow\downarrow)-\HPVTV(\downarrow\uparrow,\downarrow\uparrow)-\HPVTV(\downarrow\downarrow,\downarrow\downarrow)\}\,\label{eq:Df-PVTV}\end{align}
 with \begin{equation}
H(m_{s1}'\, m_{s2}',m_{s1}\, m_{s2})\equiv\,^{(-)}\bra{m_{s1}'\, m_{s2}'}H\ket{m_{s1}\, m_{s2}}^{(+)}\,,\end{equation}
 where $H$ is $H_{\PV}$ or $H_{\PVTV}$. The distorted (by the strong
interaction) wave functions are obtained by solving the Lippmann-Schwinger
equation\begin{equation}
\ket{m_{s1}\, m_{s2}}^{(\pm)}=\ket{m_{s1}\, m_{s2}}^{(0)}+\frac{1}{E-H_{0}-H_{\ssst{S}}\pm i\,\epsilon}\,\ket{m_{s1}\, m_{s2}}^{(\pm)}\,,\end{equation}
 where $\ket{m_{s1}\, m_{s2}}^{(0)}$ is simply a plane wave. We have
used several high-quality local $n\, p$ potentials, \emph{viz.} AV18~\cite{Wiringa:1995wb},
Reid93 and Nijm-II~\cite{Stoks:1994wp}, as input for $H_{\sst{S}}$.
The $H_{\PV}$ and $H_{\PVTV}$ used in this work are both built upon
the one-meson-exchange model and parametrized by the corresponding
$\PV$ and $\PVTV$ meson-nucleon coupling constants $h_{\ssst{M}}^{\ssst{I}}$'s
and $\bar{g}_{\ssst{M}}^{\ssst{I}}$'s ({}``$M$'' for the type
of meson and {}``$I$'' for isospin), respectively. The former is
the well-known, so-called DDH potential~\cite{Desplanques:1980hn},
which contains 6 $\PV$ couplings (with $h_{\rho}^{1\,'}$ usually
being ignored) due to one $\pi^{\pm}$-, $\rho$-, and $\omega$-exchanges,
and the most complete form of the latter, which contains 10 $\PVTV$
couplings due to one $\pi$-, $\eta$-, $\rho$- and $\omega$-exchanges,
can be found in Ref.~\cite{Liu:2004tq}. In the low-energy region,
only the lowest partial waves are important, and the results depend
on three $S$--$P$ amplitudes: $^{3}S_{1}$--$^{3}P_{1}$ ($\Delta I=1$),
$^{3}S_{1}$--$^{1}P_{1}$ ($\Delta I=0$), and $^{1}S_{0}$--$^{3}P_{0}$
($\Delta I=0,\,2$). The small admixture of $^{3}D_{1}$ to $^{3}S_{1}$
by the tensor force can be ignored safely.

The threshold behavior is examined across a wide range of neutron
energy $E_{n}$ from epithermal $\sim\textrm{eV}$ to very cold $\sim10^{-4}\,\textrm{eV}$.
Our numerical results agree very well with the qualitative predictions
by Stodolsky~\cite{Stodolsky:1981vn} that $\widetilde{\phi}_{y}$
is constant and $\widetilde{P}_{y}$ decreases as $\sqrt{E_{n}}$.
Stodolsky also pointed out that the existence of exothermic processes,
\emph{i.e.}, inelastic channels, could possibly lead to a non-zero
contribution to $\widetilde{P}_{y}$ at zero energy for neutron-nucleus
scattering. However, this is not the case for $n\, p$ scattering:
As it is known that the neutron-helicity-dependent differential cross
section for radiative capture, \emph{i.e.}, $\vec{n}+p\rightarrow d+\gamma$,
takes the form $d\,\sigma_{\pm}\propto(1\pm A_{\gamma}\,\cos\theta)$
(see, \emph{e.g.}, Ref.~\cite{Kaplan:1998xi}), the total cross sections
for neutrons of opposite helicities are the same; hence, no total
asymmetry arises from this particular exothermic process.~%
\footnote{In other words, the existence of exothermic channels is only a necessary
but not a sufficient condition for a non-zero total asymmetry at zero
energy.%
}

The target density, to which all optical observables are proportional,
is certainly an important factor affecting the feasibility of a neutron
transmission experiment. For the $\vec{n}\, p$ case, high-purity
liquid para-hydrogen, with $N\sim0.4\times10^{23}/\textrm{cm}^{3}$,
provides a good choice for the $\PV$ study~\cite{Markov:2002}.
For the $\vec{n}\,\vec{p}$ case, a target containing polarized protons
with a reasonably high density is required. A novel technique to produce
a polarized solid HD target~\cite{Honig:1995}, called SPHICE (Strongly
Polarized Hydrogen ICE), with a $95\%$ proton polarization in a molecular
volume $20\,\textrm{cm}^{3}/\textrm{mole}$, suggests that $\widetilde{N}\sim0.3\times10^{23}/\textrm{cm}^{3}$
is possible. Therefore, for the following numerical results, we adopt
$N=\widetilde{N}\sim0.4\times10^{23}/\textrm{cm}^{3}$. 

Assuming that the target density is uniform, the differential observables
$d\,\widetilde{\phi}_{y}/d\, z$ and $d\,\widetilde{P}_{y}/d\, z$
for neutrons at thermal energy, $E_{n}=0.025\,\textrm{eV}$, are given
in Tables~\ref{tab:dphi/dz} and~\ref{tab:dPz/dz}. The dominance
of pion exchange, due to its comparatively long range, is obvious.
Also its model dependence is very small. Of the three contributing
$S$--$P$ amplitudes, the $^{1}S_{0}$--$^{3}P_{0}$ transition plays
the most important role. It gives a $\bar{g}_{\pi}^{0}-4\,\bar{g}_{\pi}^{2}$
dependence on the $\PVTV$ pion-nucleon couplings. Since the heavy-meson
contributions are more sensitive to the short-range wave functions,
the difference between various strong potentials becomes more apparent.
At this energy, $d\,\widetilde{\phi}_{y}/d\, z$ is about three orders
of magnitude bigger than $d\,\widetilde{P}_{y}/d\, z$, therefore,
spin-rotation experiments look more promising, besides the advantage
already mentioned above that they are true null tests. We also calculate
$\phi_{z}$ using the same wave functions. For the AV18 model, the
result is \begin{equation}
d\,\phi_{z}/d\, z=1.130\, h_{\pi}^{1}-0.283\, h_{\rho}^{0}+0.008\, h_{\rho}^{1}+0.250\, h_{\rho}^{2}-0.269\, h_{\omega}^{0}-0.024\, h_{\omega}^{1}\,\textrm{rad/m}\,.\end{equation}
 Using the DDH {}``best values''~\cite{Desplanques:1980hn} for
the $h_{\ssst{M}}^{\ssst{I}}$'s, one gets $d\,\phi_{z}/d\, z\simeq6.5\times10^{-7}\,\textrm{rad/m}$.~%
\footnote{This number differs somewhat from a recent calculation by Schiavilla
\emph{et al.}~\cite{Schiavilla:2004wn}. This is because we use different
strong parameters and because the Yukawa function in their work is
modified by a monopole form factor. When we use the same model as
they did, we get a perfect agreement with their result.%
}

\begin{table}

\caption{$d\,\widetilde{\phi}_{y}/d\, z$ in units of $\textrm{rad/m}$ at
thermal neutron energy, $E_{n}=0.025\,\textrm{eV}$, calculated with
various strong potential models. Each entry denotes the 
multiplicative coefficient for its corresponding $\PVTV$ coupling 
constant, and the full result is the sum of every "entry$\times$coupling" 
in the same row.  
\label{tab:dphi/dz}}

\begin{tabular}{ccccccccccc}
&
 $\bar{g}_{\pi}^{0}$&
 $\bar{g}_{\pi}^{1}$&
 $\bar{g}_{\pi}^{2}$&
 $\bar{g}_{\eta}^{0}$&
 $\bar{g}_{\eta}^{1}$&
 $\bar{g}_{\rho}^{0}$&
 $\bar{g}_{\rho}^{1}$&
 $\bar{g}_{\rho}^{2}$&
 $\bar{g}_{\omega}^{0}$&
 $\bar{g}_{\omega}^{1}$\tabularnewline
\hline
AV18&
 7.758&
 1.131&
 $-$28.180&
 0.082&
 0.019&
 $-$0.046&
 0.009&
 0.157&
 $-$0.106&
 $-$0.025\tabularnewline
Reid93&
 7.735&
 1.141&
 $-$28.025&
 0.080&
 0.020&
 $-$0.046&
 0.010&
 0.152&
 $-$0.101&
 $-$0.029\tabularnewline
Nijm-II&
 7.718&
 1.153&
 $-$27.971&
 0.079&
 0.022&
 $-$0.045&
 0.011&
 0.149&
 $-$0.099&
 $-$0.033\tabularnewline
\end{tabular}
\end{table}

\begin{table}

\caption{$d\,\widetilde{P}_{y}/d\, z$ in units of $10^{-3}/\textrm{m}$ at
thermal neutron energy, $E_{n}=0.025\,\textrm{eV}$, calculated with
various strong potential models and tabulated in the same manner as 
Table\,\ref{tab:dphi/dz}. 
\label{tab:dPz/dz}}

\begin{tabular}{ccccccccccc}
&
 $\bar{g}_{\pi}^{0}$&
 $\bar{g}_{\pi}^{1}$&
 $\bar{g}_{\pi}^{2}$&
 $\bar{g}_{\eta}^{0}$&
 $\bar{g}_{\eta}^{1}$&
 $\bar{g}_{\rho}^{0}$&
 $\bar{g}_{\rho}^{1}$&
 $\bar{g}_{\rho}^{2}$&
 $\bar{g}_{\omega}^{0}$&
 $\bar{g}_{\omega}^{1}$\tabularnewline
\hline
AV18&
 2.830&
 $-$0.106&
 $-$11.589&
 0.036&
 $-$0.002&
 $-$0.016&
 $-$0.001&
 0.065&
 $-$0.047&
 0.002\tabularnewline
Reid93&
 2.814&
 $-$0.107&
 $-$11.532&
 0.035&
 $-$0.002&
 $-$0.015&
 $-$0.001&
 0.063&
 $-$0.046&
 0.003\tabularnewline
Nijm-II&
 2.808&
 $-$0.108&
 $-$11.506&
 0.035&
 $-$0.002&
 $-$0.015&
 $-$0.001&
 0.061&
 $-$0.044&
 0.003\tabularnewline
\end{tabular}
\end{table}

We now assume that a neutron spin rotation experiment with polarized
protons as target can reach a similar sensitivity of $2.7\times10^{-7}\,\textrm{rad/m}$
as what is expected for the one using para-hydrogen for the $\PV$
experiment. Our calculation then demonstrates that this null test
for $T$ violation constrains the $\PVTV$ $N\! N$ interaction at
the level of \begin{equation}
\pm2.7\times10^{-7}>7.7\,\bar{g}_{\pi}^{0}+1.1\,\bar{g}_{\pi}^{1}-28\,\bar{g}_{\pi}^{2}+\ldots\,.\label{eq:constraint-spin}\end{equation}
 On the other hand, the neutron EDM $d_{n}$ can also be expressed
in terms of these $\PVTV$ meson-nucleon couplings~\cite{Crewther:1979pi}.
By using the recent estimate in Ref.~\cite{Liu:2004tq}, the current
most stringent upper limit on $d_{n}$: $d_{n}<6.3\times10^{-26}\, e\textrm{--cm}$~\cite{Harris:1999jx},
provides the constraint\begin{equation}
\pm6.3\times10^{-11}>14\,(\bar{g}_{\pi}^{0}-\bar{g}_{\pi}^{2})+\ldots\,.\label{eq:constrain-dn}\end{equation}
 Comparing Eqs.~(\ref{eq:constraint-spin}) and~(\ref{eq:constrain-dn}),
a neutron EDM measurement at the $10^{-25}\, e\textrm{--cm}$ level
is $3$--$4$ orders of magnitude more sensitive than a spin rotation
measurement in polarized hydrogen at the $10^{-7}\,\textrm{rad/m}$
level. Given that the accuracy of $10^{-7}\,\textrm{rad/m}$ is already
state-of-the-art and that there are many difficulties involved in
keeping de-polarization effects under control, it seems very unlikely
that a neutron spin rotation experiment can compete with the neutron
EDM experiments in the near future.

However, the situation could be quite different when certain heavy
nuclei are chosen as targets. By exploiting the low-lying $p$-wave
resonances in neutron-nucleus scattering, the combined dynamical and
resonance enhancements for $\PV$ and $\PVTV$ signals could be as
large as $10^{6}$~\cite{Bunakov:1982is,Bunakov:1982kc}. A recent
$\PV$ neutron transmission measurement that exploits the $0.734\,\textrm{eV}$
$p$-wave resonance of $^{139}\textrm{La}$ resulted in $d\,\phi_{z}/d\, z=(7.4\pm1.1)\times10^{-1}\,\textrm{rad/m}$~\cite{Haseyama:2001mg}.
Compared to the theoretical prediction for thermal neutrons in hydrogen:
$d\,\phi_{z}/d\, z=5.1\textrm{--}7.2\times10^{-7}\,\textrm{rad/m}$~\cite{Schiavilla:2004wn},
one does find a $10^{6}$ enhancement factor. Therefore, if a similar
$\PVTV$ measurement could be performed with a polarized $^{139}\textrm{La}$
target and with a $10^{-7}\,\textrm{rad/m}$ sensitivity, it will
be competitive to the currently planned $d_{n}$ measurements that
target the $10^{-27}\textrm{--}10^{-28}\, e\textrm{--cm}$ level.

While this is an optimistic conclusion, there exist several major
challenges. On the experimental side, noticeably, the sensitivity
reported for $\PV$ in the $^{139}\textrm{La}$ case is only at the
$10^{-1}\,\textrm{rad/m}$ level. This six-orders-of-magnitude loss
of sensitivity thus neutralizes the $10^{6}$ enhancement factor,
which results in a measurement not better than the one with a hydrogen
target and a $10^{-7}\,\textrm{rad/m}$ sensitivity. A rough theoretical
estimate that a $3$--$4$ orders of improvement is necessary to keep
these measurements competitive to the current $d_{n}$ limit was given
in Refs.~\cite{Herczeg:1987,McKellar:1987}, and a possibility of
such an experimental improvement was reported in Ref.~\cite{Masuda:2000}.
On the theoretical side, it will require a major effort to interpret
the observables in neutron-nucleus scattering, in terms of $\PVTV$
meson-nucleon couplings, at a similar level of accuracy as what we
have done here for the $n\, p$ system ($1\%$ or even, say, at the
$10\%$ level). There have been efforts to apply the theory of statistical
spectroscopy to interpret $\PV$ phenomena (see, \emph{e.g.}, Ref.~\cite{Tomsovic:1999yb}),
apparently, how they can constrain the underlying $N\! N$ interactions
is then subject to statistics. Similar work for $T$ violation will
be necessary.

\begin{ack}
Part of this work was supported by the Dutch Stichting vor Fundamenteel
Onderzoek der Materie (FOM) under program 48 (TRI$\mu$P). We also
acknowledge support from the EU RTD network under contract HPRI-2001-50034
(NIPNET).
\end{ack}
\bibliographystyle{elsart-num-mod}
\bibliography{NeutronOptics}

\end{document}